\documentstyle[prl,aps,epsf]{revtex}
\begin{document}
\twocolumn[%
\hsize\textwidth\columnwidth\hsize\csname@twocolumnfalse\endcsname
\title{%
\hfill{\normalsize\vbox{\hbox{May 2001} \hbox{DPNU-01-09} }}\\
\vspace{-0.5cm}
\bf Fate of Vector Dominance in the Effective Field Theory}
\author{{\bf Masayasu Harada} and {\bf Koichi Yamawaki}}
\address{Department of Physics, Nagoya University,
Nagoya, 464-8602, Japan.}
\maketitle

\begin{abstract}
We reveal the full phase structure of the effective field theory for
QCD, based on the hidden local symmetry (HLS) through the one-loop
renormalization group equation including quadratic divergences.
We then show that vector dominance (VD) is not a sacred discipline of
the effective field theory but rather an accidental phenomenon
peculiar to three-flavored QCD. In particular, the chiral symmetry
restoration in HLS model takes place in a wide phase boundary surface,
on which the VD is realized nowhere. This suggests that VD may not be
valid for chiral symmetry restoration in hot and/or dense QCD.
\end{abstract}
\vskip1pc]

Since Sakurai advocated Vector Dominance (VD) as well as vector
meson universality~\cite{Sakurai},
VD has been a widely accepted notion in describing vector meson
phenomena in hadron physics. 
In fact several models such as the gauged sigma model~\cite{GG:Mei}
are based on VD to introduce 
the photon field into the Lagrangian. Moreover, it is often taken for
granted in analysing the dilepton spectra 
to probe the phase of quark-gluon plasma for the hot
and/or dense QCD~\cite{VDT}.

As far as the well-established hadron physics for the $N_f=3$ case is
concerned,
it in fact has been extremely successful in many processes such as the
electromagnetic form factor of the pion~\cite{Sakurai} and the 
electromagnetic $\pi\gamma$ transition
form factor (See, e.g., Ref.~\cite{CELLO}.), etc. However, there has
been no theoretical justification for VD 
and as it stands might be no more than a mnemonic useful only for the
three-flavored QCD at zero
temperature/density. Actually, 
{\it VD is already violated} for the
three-flavored QCD for the 
anomalous processes such as 
$\gamma \rightarrow 3 \pi/\pi^0 \rightarrow
2\gamma$~\cite{Fuj:85,BKY:88,HY:PR}
and $\omega\pi$ transition form factor (See, e.g.,
Ref.~\cite{BH:BGP}.).
This strongly suggests that VD may not be a sacred discipline of
hadron physics but 
may largely be violated in the different parameter space than the
ordinary three-flavored QCD (non-anomalous processes) such as in the
large $N_f$ QCD, $N_f$ being number of massless flavors,
and hot and/or dense QCD
where the chiral symmetry 
restoration is expected to occur. It is rather crucial whether or not
VD is still valid 
when probing such a chiral symmetry restoration  through vector meson
properties~\cite{Pisarski:2,BR}.

Here we emphasize that in the Hidden Local Symmetry (HLS)
model~\cite{BKUYY,BKY:88}
{\it the vector mesons are formulated 
precisely as gauge bosons}; nevertheless 
{\it VD as well as the universality is merely a dynamical consequence}
characterized by the parameter choice $a=2$.

In this paper we reveal the full phase structure of the effective
field theory including the vector 
mesons, based on the one-loop Renormalization Group Equation (RGE) of
HLS model.
It turns out that
{\it in view of the phase diagram
VD is very accidentally realized and only for $N_f=3$ QCD}. 
On the other hand, we find 
{\it a wide phase boundary surface of chiral
symmetry restoration} in HLS model, 
{\it on which the VD is realized nowhere}.
Furthermore,
{\it only a single point} of the phase boundary is
shown to be 
{\it selected by QCD} through the Wilsonian 
matching~\cite{HY:matching},
which actually 
coincides with the 
{\it Vector Manifestation (VM)}~\cite{HY:VM}
realized for large $N_f$ QCD
where {\it VD is badly violated with $a=1$}.

Let us first describe the HLS model based on the
$G_{\rm global} \times H_{\rm local}$ symmetry, where
$G = \mbox{SU($N_f$)}_{\rm L} \times 
\mbox{SU($N_f$)}_{\rm R}$  is the 
global chiral symmetry and 
$H = \mbox{SU($N_f$)}_{\rm V}$ is the HLS.
The basic quantities are 
the gauge bosons $\rho_\mu = \rho_\mu^a T_a$ of the HLS 
and two 
SU($N_f$)-matrix valued variables $\xi_{\rm L}$ and 
$\xi_{\rm R}$. 
They are parametrized as
$
\xi_{\rm L,R} = e^{i\sigma/F_\sigma} e^{\mp i\pi/F_\pi}
$,
where $\pi = \pi^a T_a$
denotes the pseudoscalar
Nambu-Goldstone (NG) bosons associated with 
the spontaneous breaking of $G$ and 
$\sigma = \sigma^a T_a$
the NG bosons absorbed into the HLS
gauge bosons $\rho_\mu$ which is 
identified with the vector mesons.
$F_\pi$ and $F_\sigma$ are relevant decay constants, and
the parameter $a$ is defined as
$a \equiv F_\sigma^2/F_\pi^2$.
$\xi_{\rm L}$ and $\xi_{\rm R}$ transform as
$\xi_{\rm L,R}(x) \rightarrow 
h(x) \xi_{\rm L,R}(x) g^{\dag}_{\rm L,R}$,
where $h(x) \in H_{\rm local}$ and 
$g_{\rm L,R} \in G_{\rm global}$.
The covariant derivatives of $\xi_{\rm L,R}$ are defined by
$
D_\mu \xi_{\rm L} =
\partial_\mu \xi_{\rm L} - i g \rho_\mu \xi_{\rm L}
+ i \xi_{\rm L} {\cal L}_\mu
$,
and similarly with replacement ${\rm L} \leftrightarrow {\rm R}$,
${\cal L}_\mu \leftrightarrow {\cal R}_\mu$,
where $g$ is the HLS gauge coupling, and
${\cal L}_\mu$ and ${\cal R}_\mu$ denote the external gauge fields
gauging the $G_{\rm global}$ symmetry.

The HLS Lagrangian is given by~\cite{BKUYY,BKY:88}
\begin{equation}
{\cal L} = F_\pi^2 \, \mbox{tr} 
\left[ \hat{\alpha}_{\perp\mu} \hat{\alpha}_{\perp}^\mu \right]
+ F_\sigma^2 \, \mbox{tr}
\left[ 
  \hat{\alpha}_{\parallel\mu} \hat{\alpha}_{\parallel}^\mu
\right]
+ {\cal L}_{\rm kin}(\rho_\mu) \ ,
\label{Lagrangian}
\end{equation}
where ${\cal L}_{\rm kin}(\rho_\mu)$ denotes the kinetic term of
$\rho_\mu$ 
and
\begin{eqnarray}
&&
\hat{\alpha}_{\stackrel{\perp}{\scriptscriptstyle\parallel}}^\mu =
\left( 
  D_\mu \xi_{\rm R} \cdot \xi_{\rm R}^\dag \mp 
  D_\mu \xi_{\rm L} \cdot \xi_{\rm L}^\dag
\right)
/ (2i) \ .
\end{eqnarray}
By taking the unitary gauge, $\xi_{\rm L}^\dag = \xi_{\rm R}$
($\sigma=0$), the Lagrangian in Eq.~(\ref{Lagrangian}) gives the
following tree level relations for the vector meson mass $m_\rho$, 
the $\rho$-$\gamma$ transition strength $g_\rho$, the $\rho\pi\pi$
coupling constant $g_{\rho\pi\pi}$ and the direct $\gamma\pi\pi$
coupling constant $g_{\gamma\pi\pi}$:~\cite{BKUYY,BKY:88}
\begin{eqnarray}
&& m_\rho^2 = a g^2 F_\pi^2 \ , \quad
   g_{\rho\pi\pi} = \frac{1}{2} a g \ , \nonumber\\
&& g_\rho = a g F_\pi^2 \ , \quad
   g_{\gamma\pi\pi} = \left( 1 - \frac{a}{2} \right) e \ ,
\label{physical predictions}
\end{eqnarray}
where $e$ is the electromagnetic coupling constant.

Expressions for $g_{\rho\pi\pi}$ and $g_\rho$ 
in Eq.~(\ref{physical predictions}) 
lead to the celebrated 
Kawarabayashi-Suzuki-Riazuddin-Fayyazuddin (KSRF)
relation~\cite{KSRF} (version I)
\begin{equation}
g_\rho = 2 F_\pi^2 g_{\rho\pi\pi} \ ,
\label{KSRF I}
\end{equation}
{\it independently of the parameter $a$}.
This is the low energy theorem of the HLS~\cite{BKY:85}, 
which was proved at one-loop~\cite{HY}, and then at 
any loop order~\cite{LET}.
On the other hand, making a dynamical assumption
of a parameter choice $a=2$, 
the following outstanding
phenomenological facts are reproduced from 
Eq.~(\ref{physical predictions}):~\cite{BKUYY,BKY:88}
\begin{enumerate}
\renewcommand{\labelenumi}{(\theenumi)}
\item $g_{\rho\pi\pi} = g$ 
  (universality of the $\rho$-coupling)~\cite{Sakurai}
\item $m_\rho^2 = 2 g_{\rho\pi\pi}^2 F_\pi^2$
  (KSRF II)~\cite{KSRF}
\item $g_{\gamma\pi\pi}=0$ (vector dominance of the
electromagnetic form factor of the $\pi$)~\cite{Sakurai}
\end{enumerate}
Thus, {\it even though the vector mesons are gauge bosons}
in the HLS model, {\it VD as well as the universality is not
automatic} consequence but rather dynamical one of a parameter choice
of $a=2$.

Actually, due to quantum corrections
the parameters change their values by the energy scale,
which are determined by the RGE's.
Accordingly,
values of the parameters $F_\pi$, $a$ and $g$
cannot be freely chosen,
although they are independent at tree level.
Thus, we first study the RG flows of the parameters
and the phase structure of the HLS to classify the parameter space.
Here we stress that {\it thanks to the gauge symmetry}
in the HLS model it is possible to perform
a {\it systematic loop expansion} including the vector mesons in
addition to the pseudoscalar
mesons~\cite{Georgi,HY,Tanabashi,HY:matching,HY:PR}
in a way to extend the chiral perturbation theory~\cite{ChPT}.
There the loop expansion corresponds to the
derivative expansion, so that the one-loop calculation of the RGE is
reliable in the low energy region.

As shown in Ref.~\cite{HY:letter,HY:matching}
it is important to include {\it quadratic divergences}
in calculating the quantum corrections.
Due to quadratic divergences in
the HLS dynamics, it follows that
{\it even if the bare theory defined by the cutoff $\Lambda$
is written as if it were in the broken
phase characterized by $F_\pi^2(\Lambda) > 0$},
{\it the quantum theory can be in the symmetric
phase characterized by $F_\pi^2(0)=0$}~\cite{HY:letter}.
The one-loop RGE's for $F_\pi$, $a$ and $g$
including quadratic divergences are given
by~\cite{HY:letter,HY:matching}
\begin{eqnarray}
&& \mu \frac{d F_\pi^2}{d\mu} = C
\left[ 3 a^2 g^2 F_\pi^2 + 2 (2-a) \mu^2 \right]
\ ,
\nonumber\\
&& \mu \frac{d a}{d \mu} = - C (a-1)
\left[ 3a (a+1) g^2 - (3a-1) \frac{\mu^2}{F_\pi^2} \right]
\ ,
\nonumber\\
&& \mu \frac{d g^2}{d \mu} = - C
\frac{ 87 - a^2}{6} g^4 \ .
\label{RGE Fp a g}
\end{eqnarray}
where $C = N_f/\left[2(4\pi)^2\right]$ and $\mu$ is the renormalization
scale.
It is convenient to use the following quantities:
\begin{eqnarray}
X(\mu) \equiv C \mu^2/F_\pi^2(\mu)
\ , \quad
G(\mu) \equiv C g^2(\mu) \ .
\label{def X G}
\end{eqnarray}
Then, the RGE's in Eq.~(\ref{RGE Fp a g})
are rewritten as
\begin{eqnarray}
&& \mu \frac{d X}{d\mu} = ( 2 - 3 a^2 G) X - 2 (2-a) X^2 
\ , 
\nonumber\\
&& \mu\frac{d a}{d\mu} = - (a-1) \left[ 3a (a+1) G - (3a-1) X \right]
\ ,
\nonumber
\\
&& \mu\frac{d G}{d \mu} = - \frac{ 87 - a^2}{6} G^2 
\ .
\label{RGE X a G}
\end{eqnarray}
It should be noticed that the RGE's in Eq.~(\ref{RGE X a G})
are valid above the $\rho$
mass scale $m_\rho$, where $m_\rho$ is defined by 
the on-shell condition
$m_\rho^2 = a(m_\rho) g^2(m_\rho) F_\pi^2(m_\rho)$.
In terms of $X$, $a$ and $G$, the on-shell condition becomes
$a(m_\rho) G(m_\rho) = X(m_\rho)$.
Then the region where the RGE's in Eq.~(\ref{RGE X a G})
are valid is specified by the condition
$a(\mu) G(\mu) \le X(\mu)$.

Seeking the parameters for which all right-hand-sides of three RGE's
in Eq.~(\ref{RGE X a G}) vanish {\it simultaneously},
we obtain {\it three fixed points} and {\it one fixed line
in the physical region} and one fixed point in the
unphysical region (i.e., $a<0$ and $X<0$).
Those in the physical region (labeled by $i=1,\ldots,4$)
are given by 
\begin{eqnarray}
&& \left(X^\ast_i,\, a^\ast_i,\, G^\ast_i\right)
  = \left( 0,\, \mbox{any},\, 0 \right) \ , 
  \quad
  \left( 1,\, 1,\, 0 \right) \ ,
  \quad
  \left( \frac{3}{5},\, \frac{1}{3},\, 0 \right) \ ,
\nonumber\\
&& \quad
  \left( 
    \frac{2(2+45\sqrt{87})}{4097},\, 
    \sqrt{87},\, 
    \frac{2(11919-176\sqrt{87})}{1069317}
  \right)
\ .
\label{fixed points}
\end{eqnarray}

\begin{figure}[tbhp]
\begin{center}
\epsfxsize = 6.5cm
\ \epsfbox{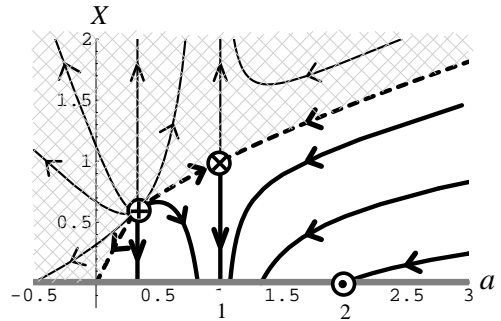}
\end{center}
\caption[]{Phase diagram on $G=0$ plane.
Arrows on the flows
are written from the ultraviolet to the infrared.
Gray line denotes the
fixed line $( X^\ast_1, a^\ast_1, G^\ast_1 ) = 
( 0, \mbox{any}, 0 )$.
Points indicated by $\oplus$ and $\otimes$ (VM point)
denote
the fixed points
$( 3/5, 1/3, 0 )$ and 
$( 1, 1, 0 )$, respectively.
Dashed lines divide the broken phase (lower side) and the
symmetric phase (upper side; cross-hatched area):
Flows drawn by thick lines are in the broken phase,
while those by thin lines are in the symmetric phase.
The point indicated by $\odot$,
$( X, a, G ) = ( 0, 2, 0 )$,
correspond to the VD, $a(0)=2$.
}\label{fig:flows G0}
\end{figure}

Note that $G=0$ is a fixed point of the RGE for $G$, and $a=1$ is the
one for $a$.
Hence RG flows on $G=0$ plane and $a=1$ plane are confined in the
respective planes.

Let us first study the phase structure of the HLS for $G=0$
(see Fig.~\ref{fig:flows G0})
in which case $m_\rho$ vanishes and the RGE's (\ref{RGE X a G}) are
valid all the way down to the low energy limit, $\mu \ge m_\rho = 0$.
There are one fixed line and two fixed points
[first three in Eq.~(\ref{fixed points})].
Generally,
the phase boundary is specified by $F_\pi^2(0)=0$,
namely, governed by the infrared fixed point such that 
$X(0)\neq0$ (see Eq.~(\ref{def X G})).
Such a fixed point is the point
$( X^\ast_2,a^\ast_2,G^\ast_2 ) = (1,1,0)$,
which is nothing but the VM point~\cite{HY:VM}.
Then the phase boundary is given by 
the RG flows entering 
$( X^\ast_2,a^\ast_2,G^\ast_2 )$.
Since $a=1/3$ is a fixed point of the RGE for $a$ in 
Eq.~(\ref{RGE X a G}),
the RG flows for $a<1/3$ cannot enter 
$( X^\ast_2,a^\ast_2,G^\ast_2 )$.
Hence there is no phase boundary specified by 
$F_\pi^2(0)=0$ in $a<1/3$ region.
Instead,
$F_\sigma^2(0)$ vanishes even though $F_\pi^2(0)\neq0$,
namely $a(0)=X(0)=0$.
Then the phase boundary for $a<1/3$
is given by the RG flow entering the point
$(X,a,G) = (0,0,0)$.
In Fig.~\ref{fig:flows G0}
the phase boundary is drawn by the dashed line,
which divides the phases into 
the symmetric phase~\cite{foot:sym}
(upper side; cross-hatched area)
and the broken one (lower side).

\begin{figure}[tbhp]
\begin{center}
\epsfxsize = 6.5cm
\ \epsfbox{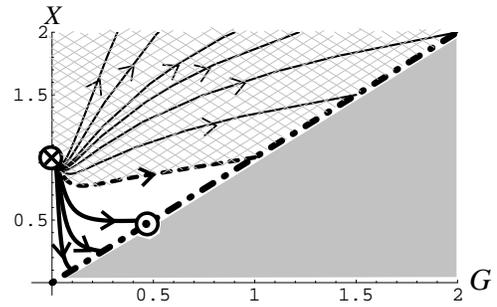}\\
\end{center}
\caption[]{Phase diagram on $a=1$ plane.
Arrows on the flows
are written from the ultraviolet to the infrared.
Point indicated by $\otimes$ denotes
the VM fixed point 
$(X^\ast_2,a^\ast_2,G^\ast_2) = \left( 1, 1, 0 \right)$.
Flows drawn by thick lines are in the broken phase,
while those by thin lines are in the symmetric phase
(cross-hatched area).
Dot-dashed line corresponds to the on-shell condition $G=X$.
In the shaded area the RGE's (\ref{RGE X a G}) are not valid
since $\rho$ has already decoupled.
Point indicated by $\odot$,
$\left( 1/2, 1, 1/2 \right)$,
corresponds to the VD, $a(0)=2$. (See Eq.~(\ref{def:a0}).)
}\label{fig:flows a1}
\end{figure}

In the case of $G>0$, on the other hand,
the $\rho$ becomes massive ($m_\rho\neq 0$),
and thus decouples at $m_\rho$ scale.
Below the $m_\rho$ scale
$a$ and $G$ no longer run, while $F_\pi$ still runs 
by the $\pi$ loop effect.
Thus, to study the phase structure for $G>0$
we need the RGE for $F_\pi$
for $\mu < m_\rho$ (denoted by $F_\pi^{(\pi)}$).
This
is given by
$d[F_\pi^{(\pi)}]^2/d\mu^2 = 2C$~\cite{HY:letter},
which is readily solved as
\begin{equation}
\left[F_\pi^{(\pi)}(\mu)\right]^2 = 
\left[F_\pi^{(\pi)}(m_\rho)\right]^2
- 2 C \left( m_\rho^2 - \mu^2 \right) \ .
\label{sol Fp pi}
\end{equation}
Then {\it the quadratic divergence} 
(second term in Eq.~(\ref{sol Fp pi}))
{\it of the $\pi$ loop can give rise to chiral symmetry restoration}
$F_\pi^{(\pi)}(0) = 0$~\cite{HY:letter}.
Thus the phase boundary is specified by the condition
$[F_\pi^{(\pi)}(m_\rho)]^2 = 2 C m_\rho^2$.
Note that
the relation between $[F_\pi^{(\pi)}(m_\rho)]^2$
and $F_\pi^2(m_\rho)$ including the finite renormalization effect
is given by~\cite{HY:matching}
\begin{equation}
\left[F_\pi^{(\pi)}(m_\rho)\right]^2 =
F_\pi^2(m_\rho) + C \, a(m_\rho) m_\rho^2 \ ,
\label{rel}
\end{equation}
which is converted into the condition for 
$X(m_\rho)$ and $a(m_\rho)$.
Combination of this with the on-shell condition
specifies the phase boundary in the full $(X,a,G)$ space,
which is given by the collection of the
RG flows entering points on the line specified by
\begin{eqnarray}
&& 2-a(m_\rho) = 1/ X(m_\rho) \ ,
\nonumber\\
&& a(m_\rho) G(m_\rho) = X(m_\rho) \ .
\label{phase boundary}
\end{eqnarray}
Such a surface can be imagined from 
Figs.~\ref{fig:flows G0} and \ref{fig:flows a1}.

We now study the $a=1$ plane
(see Fig.~\ref{fig:flows a1}).
The flows stop at the on-shell
of $\rho$ ($G=X$; dot-dashed line in Fig.~\ref{fig:flows a1})
and should be switched over to RGE of $F_\pi^{(\pi)}(\mu)$ as
mentioned above.
{}From Eq.~(\ref{phase boundary}) with $a=1$
the flow entering $(X,G) = (1,1)$
(dashed line) is the phase boundary
which distinguishes the broken phase (lower side) from the
symmetric one (upper side; cross-hatched area).

\begin{figure}[htbp]
\begin{center}
\epsfxsize=7cm
\ \epsfbox{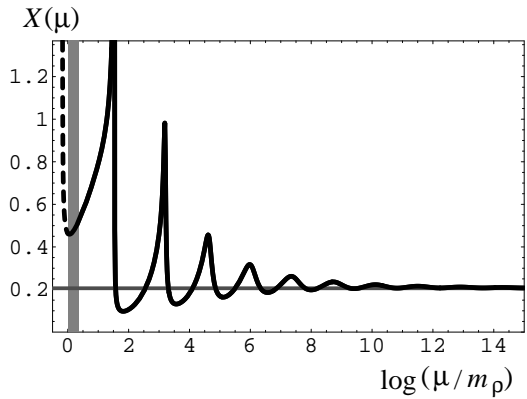}\\
\end{center}
\caption[]{Scale dependence of $X(\mu)$ in QCD with $N_f=3$.
Shaded area denotes the physical region, 
$m_\rho \le \mu \le \Lambda$.
Flow shown by the dashed line are obtained by
extending it to the (unphysical) infrared region by taking literally
the RGE's in Eq.~(\ref{RGE X a G}).
In an idealized high energy limit
the flow approaches to the fixed point
$X^\ast_4 =2(2+45\sqrt{87})/4097 \simeq 0.2$.
}\label{fig:running}
\end{figure}

For $a<1$, RG flows approach to the fixed point
$\left(X^\ast_3,a^\ast_3,G^\ast_3\right)=(3/5, 1/3, 0)$
in the idealized high energy limit ($\mu\rightarrow\infty$).

For $a>1$, RG flows in the broken phase approach to 
$\left(X^\ast_4,a^\ast_4,G^\ast_4\right)\simeq(0.2, 9.3, 0.02)$,
which is precisely the fixed point 
that {\it the RG flow of the $N_f=3$ QCD belongs to}.
To see how the RG flow of $N_f=3$ QCD
approaches to this fixed point,
we show the $\mu$-dependence of $X(\mu)$ in Fig.~\ref{fig:running}
where values of the parameters at $\mu = m_\rho$ are set to be 
$\left( X(m_\rho), a(m_\rho), G(m_\rho)\right)
\simeq \left( 0.46,  1.22, 0.38 \right)$ through Wilsonian
matching with the underlying QCD~\cite{HY:matching}.
The values of $X$ close to $1/2$
in the physical region ($m_\rho \le \mu \le \Lambda$) are
{\it very unstable against RGE flow, and hence
$X \sim 1/2$ is realized in a very accidental way}.

Let us now discuss the VD which is characterized by $a(0) = 2$.
Since $F_\sigma^2$ does not run for $\mu< m_\rho$
while $F_\pi^2$ does, we have~\cite{HY:matching}
\begin{equation}
a(\mu)
\equiv
\left\{\begin{array}{ll}
 F_\sigma^2(\mu)/F_\pi^2(\mu)
& (\mu > m_\rho ) \ ,
\\
 F_\sigma^2(m_\rho)/\left[ F_\pi^{(\pi)}(\mu) \right]^2 
& (\mu < m_\rho ) \ .
\end{array}\right.
\end{equation}
Then by using Eqs.~(\ref{sol Fp pi}) and (\ref{rel}), 
$a(0)$ is given by
\begin{equation}
a(0)
=
a(m_\rho) / \left[ 1 + a(m_\rho) X(m_\rho) - 2X(m_\rho) \right]
\ .
\label{def:a0}
\end{equation}
This implies that the VD ($a(0) =2$) is only realized
for $( X(m_\rho), a(m_\rho) ) = 
( 1/2, \mbox{any} )$ or
$( \mbox{any}, 2 )$.

In $N_f=3$ QCD, the parameters at $m_\rho$ scale,
$\left( X(m_\rho), a(m_\rho), G(m_\rho)\right)
\simeq \left( 0.46,  1.22, 0.38 \right)$,
happen to be near such a VD point.
However, the RG flow actually belongs to the fixed point
$\left(X^\ast_4,a^\ast_4,G^\ast_4\right)$ which is far away from the
VD value.
Thus, {\it the VD in $N_f=3$ QCD is accidentally realized by 
$X(m_\rho)\sim1/2$ which is very unstable against the RG flow}
(see Fig.~\ref{fig:running}).
For $G=0$ (Fig.~\ref{fig:flows G0})
the VD holds only if the parameters are
(accidentally)
chosen to be on the RG flow entering
$\left( X, a, G \right) = \left( 0, 2, 0 \right)$
(indicated by $\odot$) which is an end point of the line
$\left( X(m_\rho), a(m_\rho) \right)=(\mbox{any},2)$.
For $a=1$ (Fig.~\ref{fig:flows a1}), 
on the other hand,
the VD point
$\left( X, a, G \right) = \left( 1/2, 1, 1/2 \right)$
(indicated by $\odot$) lies on the line
$\left( X(m_\rho), a(m_\rho) \right)=(1/2,\mbox{any})$.

Then, phase diagrams in Figs.~\ref{fig:flows G0} and 
\ref{fig:flows a1} 
and their extensions to the entire parameter space
(including Fig.~\ref{fig:running})
show that neither 
$X(m_\rho) = 1/2$ nor $a(m_\rho) = 2$ is a special point in the
parameter space of the HLS.
Thus we conclude that
the {\it VD as well as the universality
can be satisfied only accidentally}.
Therefore,
when we change the parameter of QCD,
the VD is generally violated.
In particular, neither 
$X(m_\rho) = 1/2$ nor $a(m_\rho) = 2$ is satisfied on the phase
boundary surface characterized by Eq.~(\ref{phase boundary}) where the
chiral restoration takes place in HLS model.
Therefore,
{\it VD is realized nowhere on the chiral restoration surface !}

Moreover, 
{\it when the HLS is matched with QCD},
only the point
$( X^\ast_2, a^\ast_2, G^\ast_2 ) = ( 1, 1, 0 )$, 
the {\it VM point},
on the phase boundary is selected,
since 
the axialvector and vector current correlators in HLS
can be matched with those in QCD only at that point~\cite{HY:VM}.
{\it
Therefore, QCD predicts $a(0)=1$,
i.e., large violation of the VD at chiral restoration}.
Actually, for the chiral restoration
{\it in the large $N_f$ QCD}~\cite{lattice,ATW}
{\it the VM can in fact
takes place}~\cite{HY:VM}, {\it and thus the VD is badly violated}.

Finally, we suggest that
if the VM takes place in other chiral restoration such as the one in
the hot and/or dense QCD, 
the VD should be largely
violated near the critical point.

This work is supported in part by Grant-in-Aid for Scientific Research
(B)\#11695030 (K.Y.), (A)\#12014206 (K.Y.) and (A)\#12740144 (M.H.).


\begin{thebibliography}{99}
\bibitem{foot:sym}
Here ``symmetric phase'' means that
$F_\pi^2(\mu) = 0$ or $F_\sigma^2(\mu) =0$,
namely $1/X(\mu)= F_\pi^2(\mu)/C\mu^2 = 0$ or $a(\mu)=0$ 
for non-zero (finite) $\mu$.
\end{thebibliography}
\end{document}